\documentclass{article}
\usepackage{spconf,amsmath,graphicx}
\usepackage{tabularx}
\usepackage{color}
\usepackage{booktabs}
\usepackage{multirow}
\usepackage{tikz}
\usepackage{cite}
\usepackage{amssymb}

\title{On using the UA-Speech and TORGO databases to validate \\ automatic dysarthric speech classification approaches}

\name{Guilherme Schu$^{\star, \dagger}$, Parvaneh Janbakhshi$^{\ddagger}$, Ina Kodrasi$^{\star}$\thanks{This work was supported by the Swiss National Science Foundation project no CRSII5\_202228 on ``Characterisation of motor speech disorders and processes''.}}
\address{$^{\star}$Idiap Research Institute, Martigny, Switzerland \\
  $^{\dagger}$\'Ecole Polytechnique F\'ed\'erale de Lausanne, Lausanne, Switzerland\\$^{\ddagger}$ Bayer AG, Berlin, Germany\\
  \tt guilherme.garcia@idiap.ch\\
}

\def\ninept{\def\baselinestretch{.939}\let\normalsize\small\normalsize}
\begin{document}
\ninept
\maketitle
\begin{abstract}
Although the UA-Speech and TORGO databases of control and dysarthric speech are invaluable resources made available to the research community with the objective of developing robust automatic speech recognition systems, they have also been used to validate a considerable number of automatic dysarthric speech classification approaches. Such approaches typically rely on the underlying assumption that recordings from control and dysarthric speakers are collected in the same noiseless environment using the same recording setup. In this paper, we show that this assumption is violated for the UA-Speech and TORGO databases. Using voice activity detection to extract speech and non-speech segments, we show that the majority of state-of-the-art dysarthria classification approaches achieve the same or a considerably better performance when using the non-speech segments of these databases than when using the speech segments. These results demonstrate that such approaches trained and validated on the UA-Speech and TORGO databases are potentially learning characteristics of the recording environment or setup rather than dysarthric speech characteristics. We hope that these results raise awareness in the research community about the importance of the quality of recordings when developing and evaluating automatic dysarthria classification approaches.
\end{abstract}

\section{Introduction}

Dysarthria is a motor speech disorder that occurs due to brain trauma or neurological conditions such as Cerebral Palsy~(CP), Amyotrophic Lateral Sclerosis~(ALS), or Parkinson’s disease, and may affect the overall communicative ability of a patient~\cite{darley1969differential,yunusova2008articulatory}. To diagnose and manage it, speech pathologists perform auditory-perceptual assessments to evaluate different components of the speech production mechanism. 
However, these assessments can be time-consuming and subjective~\cite{kent1996hearing,connaghan2021exploratory}. 
Aiming at assisting healthcare professionals, there has been a growing interest in the research community to develop automatic dysarthria classification approaches.

State-of-the-art automatic dysarthria classification approaches can be broadly grouped into two categories, i.e., i)~approaches which use handcrafted features with classical machine learning classifiers~\cite{kadi2016fully,gillespie2017cross,narendra2018dysarthric,jeancolas2019comparison,kodrasi2020automatic,kodrasi2020spectro,janbakhshi2020subspace} and ii) deep learning approaches which are trained to automatically extract and classify discriminative speech representations~\cite{vasquez2017convolutional,millet2019learning,janbakhshi2021automatic,qi2021speech,wang2021unsupervised,gupta2021residual,janbakhshi2021supervised,vasquez2021modeling,bhattacharjee2021source,joshy2022automated,janbakhshi2022experimental}.
Commonly used approaches in the first category exploit support vector machines (SVMs) with Mel-frequency cepstral coefficients (MFCCs)~\cite{kadi2016fully}, glotal-based features~\cite{gillespie2017cross}, openSMILE features~\cite{narendra2018dysarthric}, or sparsity-based features~\cite{kodrasi2020spectro}.
In addition to SVMs, other classical machine learning methods such as Gaussian Mixture Models~\cite{jeancolas2019comparison} and subspace-based learning~\cite{janbakhshi2020subspace} have been explored.
Approaches in the second category have focused on exploring various network architectures and training paradigms such as long short-term memory networks~\cite{millet2019learning,joshy2022automated}, variational autoencoders~\cite{qi2021speech}, adversarial training~\cite{wang2021unsupervised}, and convolutional neural networks~(CNNs)~\cite{vasquez2017convolutional,janbakhshi2021automatic,bhattacharjee2021source,gupta2021residual,janbakhshi2022experimental}. 
More recently, self-supervised learning (SSL) methods such as wav2vec2~\cite{baevski2020wav2vec} have been successfully exploited for a variety of speech classification tasks~\cite{yang2021superb}, motivating their use for automatic dysarthria classification.
    
Despite the reported success of automatic dysarthria classification approaches, state-of-the-art literature typically relies on the underlying assumption that recordings from control and dysarthric speakers are obtained in the same noiseless environment using the same recording setup.
If recordings for one group of speakers are obtained in a consistently different environment than recordings for the other group of speakers, classifiers trained on such recordings would potentially learn characteristics of the recording environment instead of dysarthric speech characteristics. 
Unfortunately, such an assumption does not seem to be fulfilled for the commonly used UA-Speech~\cite{kim2008dysarthric} and TORGO~\cite{rudzicz2012torgo} databases. 
Although these databases are made available to the community to develop automatic speech recognition~(ASR) systems~(where different recording environments and setups can even be desirable in order to develop robust ASR systems), they have also been used to validate a considerable number of state-of-the-art automatic dysarthria classification approaches, such as e.g.~\cite{gillespie2017cross,narendra2018dysarthric,kadi2016fully,millet2019learning,qi2021speech,wang2021unsupervised,gupta2021residual,joshy2022automated}.

In this paper, we investigate the use of the UA-Speech and TORGO databases to validate automatic dysarthria classification approaches. 
We hypothesize that the reported classification results using these databases may be reflecting characteristics of the recording environment rather than characteristics of dysarthric speech.
To investigate this hypothesis, we first estimate the utterance-level signal-to-noise ratio (SNR) in these databases, confirming the large variability in recording conditions. 
Further, using voice activity detection~(VAD), segments that contain only speech and segments that do not contain any speech are extracted from each utterance in these databases. 
State-of-the-art dysarthria classification approaches are then trained and validated using only the speech segments or using only the non-speech segments.  
Remarkably, experimental results show that for both databases, the majority of the considered state-of-the-art approaches achieve the same or even a considerably better dysarthria classification performance when using only non-speech segments than when using only speech segments. 
The presented results confirm the hypothesis that dysarthria classification approaches validated on the UA-Speech and TORGO databases may be learning characteristics of the recording environment rather than dysarthria characteristics.

\section{UA-Speech and TORGO databases}
In the following, the UA-Speech and TORGO databases are briefly described.
Using forced alignment from ASR systems from~\cite{hermann2021handling} as VAD, speech segments and non-speech segments are extracted from each utterance in these databases. 

{\emph{UA-Speech~\cite{kim2008dysarthric}}.} \enspace The UA-Speech database contains recordings of $15$ patients with CP ($4$ females, $11$ males) and $13$ control speakers ($4$ females, $9$ males). 
Speech signals are sampled at $16$~kHz. 
Since a $7$-channel microphone array is used for recording the speakers, we consider the recordings of the $5$th-channel (arbitrarily selected) for the evaluations presented in this paper.
The number of utterances per speaker is $721$ and the average length of all utterances considered for each speaker is $1887$~s. 
Further, the average length of all extracted speech and non-speech segments for each speaker is $564$~s and $1323$~s respectively.

\begin{figure}[t!]
    \centering
    \includegraphics[width=0.81\linewidth]{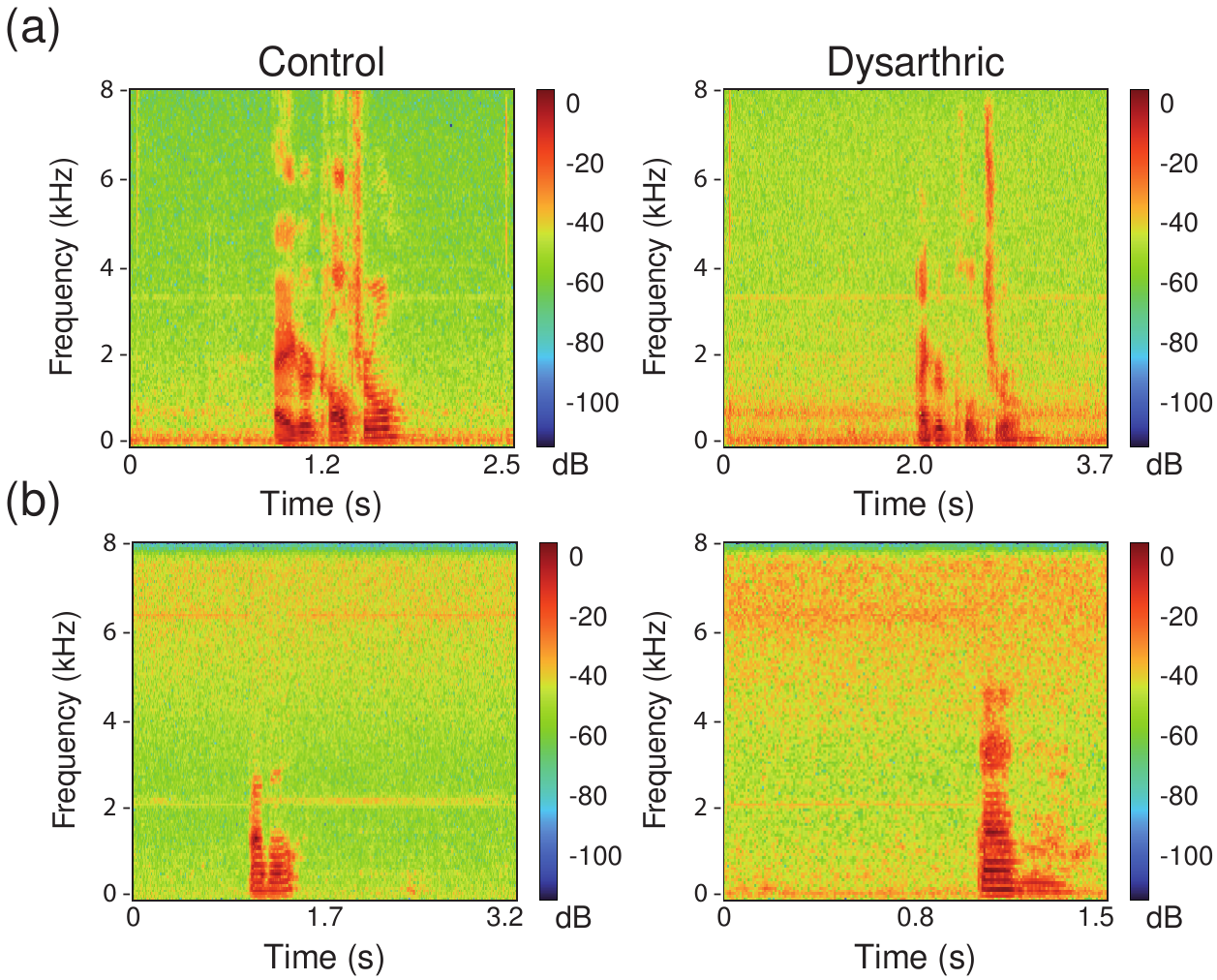}
        \vspace{-0.25cm}
    \caption{Spectrograms of an exemplary utterance from a control and dysarthric speaker from the a) UA-Speech and b) TORGO databases.}
    \label{fig:stft_noise}
\end{figure}

{\emph{TORGO~\cite{rudzicz2012torgo}.}} \enspace The TORGO database contains recordings from 7 patients ($3$ females, $4$ males) with CP or ALS and from $7$ control speakers ($3$ females, $4$ males). 
Speech signals are sampled at $16$~kHz. 
To avoid additional sources of variability besides dysarthria characteristics, we use only utterances with matched phonetic content across all speakers. 
The number of such utterances per speaker is $62$ and the average length of all utterances considered for each speaker is $201$~s. 
Further, the average length of all extracted speech and non-speech segments for each speaker is $76$~s and $125$~s respectively.

Fig.~\ref{fig:stft_noise} depicts spectrograms of exemplary utterances from a control and dysarthric speaker from the UA-Speech and TORGO databases.
Visually inspecting the spectrograms in Fig.~\ref{fig:stft_noise}(a) reveals that the exemplary dysarthric spectrogram from the UA-Speech database is noisier than the control spectrogram, showing higher levels of noise, particularly at lower frequencies.
Further, visually inspecting the spectrograms in Fig.~\ref{fig:stft_noise}(b) reveals that also the exemplary dysarthric spectrogram from the TORGO database is noisier than the control spectrogram, showing higher levels of noise, particularly at higher frequencies.

\section{Methods}
In this section, the utterance-level SNR estimator is first briefly described.
Further, details on the considered state-of-the-art dysarthria classification approaches are presented.

\subsection{SNR estimation}
Although robust SNR estimation remains an open problem, in this paper we use the recently proposed data-driven recurrent neural network from~\cite{Li_ITASLP_2021}, since it was shown to outperform several state-of-the-art SNR estimators.
We use the same network architecture, training procedure, and training and validation datasets as in~\cite{Li_ITASLP_2021}.
The input to the network is the magnitude spectrogram of the noisy signals, whereas the target of the network is the frame-level SNR.
Once the frame-level SNR is estimated, an estimate of the utterance-level SNR is obtained as in~\cite{Li_ITASLP_2021}.

\subsection{Dysarthria classification approaches}
The considered state-of-the-art dysarthria classification approaches are summarized in Table~\ref{tbl:summary}.
In the following, the handcrafted features or input representations and the classifiers used in these approaches are introduced. 
\begin{table*}[t]
\scriptsize
\centering
\renewcommand{\arraystretch}{1.3}
\caption{Summary of the investigated dysarthria classification approaches.}
\label{tbl:summary}
\begin{tabularx}{\linewidth}{XXX}
\toprule
Approach & Classifier            & Handcrafed feature or input representation                   \\
\midrule
\emph{SVM+openSMILE}~\cite{narendra2018dysarthric}     & \multirow{3}{*}{Support vector machine with RBF kernel} & ComParE-2016 - openSMILE features \\
\emph{SVM+MFCCs}~\cite{kadi2016fully}     &                       & Mel-frequency cepstral coefficients \\
\emph{SVM+sparsity-based features}~\cite{kodrasi2020spectro}     &                       & Sparsity characterized by the shape parameter \\
\midrule
\emph{CNN+Mel spectrograms}~\cite{vasquez2017convolutional}     & Convolutional neural network                   & \multirow{2}{*}{Mel spectrograms}     \\
\emph{SRL+Mel spectrograms}~\cite{janbakhshi2021supervised}     & Speech representation learning                  &                         \\
\midrule
\emph{MLP+ft-wav2vec2}~\cite{yang2021superb}     & \multirow{2}{*}{Linear classifier} & Fine-tuned wav2vec2 embeddings                      \\
\emph{MLP+wav2vec2}~\cite{yang2021superb}     &                       & Wav2vec2 embeddings without fine-tuning                       \\
\bottomrule
\end{tabularx}
\end{table*}

\subsubsection{Handcrafted features and input representations}\label{subsec: inputs}
{\emph{OpenSMILE}.} \enspace 
As in~\cite{narendra2018dysarthric,kodrasi2020automatic}, for each utterance, we extract $6373$ features used in the ComParE-$2016$ challenge~\cite{schuller2016interspeech} with the openSMILE toolkit~\cite{eyben2010opensmile}. 
Similarly to~\cite{kodrasi2020automatic}, dimensionality reduction with Principal Component Analysis is performed by selecting the number of features explaining $95\%$ of the variance in the data. 

{\emph{MFCCs.}}
We extract the mean, variance, skewness, and kurtosis of the first $12$ MFCCs coefficients using the OpenSMILE toolkit~\cite{eyben2010opensmile}, constructing a $48$-dimensional feature vector for each utterance.

{\emph{Sparsity-based features.}} \enspace 
As in~\cite{kodrasi2020spectro}, we compute sparsity-based features through the shape parameter of a Chi distribution. 
To this end, the short-time Fourier transform (STFT) of each utterance is first computed using a Hamming window of length $16$~ms and a frame shift of $8$~ms.
For each frequency bin, a maximum likelihood estimate of the shape parameter of the Chi distribution best modeling the spectral magnitude is obtained. 
At a sampling frequency of $16$~kHz, this procedure yields a $129$-dimensional feature vector for each utterance.

{\emph{Mel spectrograms.}} \enspace
Similarly to~\cite{janbakhshi2021supervised}, 
Mel-scale representations are computed for $500$~ms long segments extracted from utterances using a time shift of $250$~ms. 
For each segment, the STFT with a Hamming window of length $32$~ms and a frame shift of $4$~ms is computed. Final representations are obtained by transforming the STFT coefficients to Mel-scale using $126$ Mel bands.

{\emph{Wav2vec2.}} \enspace Wav2vec2 is a state-of-the-art SSL method that can produce powerful latent speech representations directly from the raw speech signal. 
The release of the SUPERB benchmark~\cite{yang2021superb} has demonstrated that state-of-the-art results on several speech processing tasks can be achieved by fine-tuning the wav2vec2 model with a lightweight linear prediction classifier.
Motivated by these results, in this paper we also exploit the wav2vec2 model for automatic dysarthria classification.

\subsubsection{Classifiers}
\label{subsec: methods}

{\emph{Support vector machines.}} \enspace SVMs are traditional classifiers commonly used with handcrafted acoustic features for dysarthria classification~\cite{kadi2016fully,gillespie2017cross,narendra2018dysarthric,kodrasi2020spectro}.
In the following, SVMs with a radial basis kernel function~(RBF) are used with different handcrafted acoustic features, i.e., openSMILE, MFCCs, and sparsity-based features.

{\emph{Convolutional neural networks.}}  \enspace CNNs have been widely used to extract discriminative input representations and achieve dysarthric speech classification~\cite{vasquez2017convolutional,janbakhshi2021automatic,bhattacharjee2021source,gupta2021residual,janbakhshi2022experimental}. 
In the following, we use a CNN operating on Mel-scale input spectrograms~\cite{vasquez2017convolutional}. 
We adopt the architecture from~\cite{vasquez2017convolutional} consisting of two convolutional layers (with $32$ and $64$ channels, kernel size: $10\times10$, and stride: $1$). Each convolutional layer is followed by batch normalization, max-pooling (kernel size: $2$, stride: $3$), and the ReLU activation function. 
A dropout layer with a rate of $20\%$ is placed after the final convolutional layer. 
A final fully-connected layer with $128$ input units and $2$ output units is used for dysarthria classification.

{\emph{Speech representation learning~(SRL)}.}~\enspace 
In this paper, SRL is used to refer to the state-of-the-art dysarthria classification approach proposed in~\cite{janbakhshi2021supervised}, where a CNN-based auto-encoder is used to learn low dimensional discriminative bottleneck representations from Mel-scale input spectrograms. 
Bottleneck representations are learned by jointly minimizing the auto-encoder loss and the loss of a linear dysarthria classifier.
The learned representations are then fine-tuned for the final dysarthria classification network.
The architecture description of the network can be found in~\cite{janbakhshi2021supervised}. 

{\emph{Multilayer perceptron (MLP).}}~\enspace 
Motivated by~\cite{yang2021superb}, we also evaluate the performance of an MLP trained on wav2vec2 embeddings for dysarthria classification.
The MLP consists of two fully-connected layers. 
The first layer has $768$ input units and $256$ output units and the second layer has $256$ input units and $2$ output units.
Similarly to the speaker identification task in~\cite{yang2021superb}, each utterance is processed by the wav2vec2 model and the obtained embeddings are mean-pooled prior to being forwarded to the MLP classifier. 
As outlined in Table~\ref{tbl:summary} and as described in Section~\ref{sec: tr}, we consider two approaches using wav2vec2, i.e., \emph{MLP+ft-wav2vec2} referring to fine-tuning parts of the wav2vec2 model together with the MLP for dysarthria classification and {\emph{MLP+wav2vec2} referring to freezing the wav2vec2 model and training only the MLP.

\section{Experimental Results}
\label{sec: exp}
In this section, the utterance-level SNR of control and dysarthric recordings from the UA-Speech and TORGO databases is analyzed.
Further, the performance of state-of-the-art dysarthria classification approaches when using only speech segments and only non-speech segments from these databases is analyzed. 
For completeness, the performance of the considered classification approaches when using the complete utterances without any VAD~(i.e., both speech and non-speech segments) is also presented.

\subsection{Training and validation}
\label{sec: tr}
For all approaches investigated in this paper (cf. Table~\ref{tbl:summary}), a  leave-one-speaker-out validation strategy is used.  
In each fold, $90\%$ of the data from the training speakers is used for training, whereas $10\%$ of the data is used for validation. 
The prediction for a test speaker is made through majority voting of the utterance-level/segment-level predictions and the final performance is evaluated in terms of the speaker-level classification accuracy. 
To reduce the impact that random initialization has on the final performance, we have trained all approaches using $3$ different random initialization.
The reported final performance for all approaches is the mean and standard deviation of the speaker-level classification accuracy obtained across these different models.
Except for the wav2vec2 embeddings, we apply z-score standardization to all handcrafted acoustic features and input representations.
In the following, details on the training of each considered approach are presented.

\emph{SVMs}.~\enspace Separate SVMs are trained for each handcrafted acoustic feature in Table~\ref{tbl:summary}. 
The soft margin constant $C$ and the kernel width $\gamma$ are optimized using a grid search procedure with $C \in \{10, 10^4\}$ and $\gamma \in \{10^{-4}, 10^{-1}\}$. The optimal hyperparameters are selected as the ones that achieve the highest utterance-level classification accuracy on the validation set. 

\emph{CNN+Mel spectrograms}.~\enspace
The CNN is trained using the Adam optimizer and the cross–entropy loss function.
We use a batch size of $128$ and an initial learning rate of $2 \times 10^{-5}$ for a total of $50$ epochs. 
A scheduler is set to halve the learning rate if the loss on the validation set does not decrease for $5$ consecutive iterations. 

\emph{SRL+Mel spectrograms}.~\enspace As in~\cite{janbakhshi2021supervised}, the stochastic gradient descent algorithm is used for training the SRL approach. 
The dysarthria classifier is trained using cross-entropy, whereas the auto-encoder is trained using mean square error. Further, we use a batch size of $128$ and an initial learning rate of $0.02$ for a total of $20$ epochs. A scheduler is set to halve the learning rate if the loss on the validation set does not decrease for $5$ consecutive iterations. 

\emph{MLP+ft-wav2vec2.}~\enspace 
To fine-tune the wav2vec2 model, we freeze the CNN encoder and fine-tune the transformer and the MLP classifier.
As in~\cite{yang2021superb}, the AdamW optimizer and the cross-entropy loss function are used.
Training is done with an effective batch size of $128$, i.e., a batch size of $16$ and a gradient accumulation step of $8$. A linear warm-up scheduler is used (warm-up ratio: $0.1$) and the initial learning rate is set to $3\times10^{-5}$.

\emph{MLP+wav2vec2.} \enspace Using the wav2vec2 embeddings without fine-tuning refers to freezing the complete wav2vec2 model and training only the MLP classifier for dysarthria classification.
The used optimizer, loss function, batch size, and learning rate are the same as for the \emph{MLP+ft-wav2vec2} approach.

\begin{table}[t!]
    \scriptsize
  \begin{center}
    \caption{Mean and standard deviation of the estimated SNRs [dB] across all utterances of control and dysarthric speakers in the UA-Speech and TORGO databases.}
    \label{tbl:snr}
    \begin{tabularx}{\linewidth}{Xll}
      \toprule
       Speakers & UA-Speech & TORGO \\
      \midrule
        Control & ${\color{white}{-}}3.7 \pm 11.5$ & ${\color{white}{-}}2.1 \pm 13.2$ \\
      Dysarthric & $-7.6 \pm 16.1$ & $-4.0 \pm 14.7$ \\  
      \bottomrule
   \end{tabularx}
 \end{center}
\end{table}

\subsection{Results}

{\emph{SNR estimation.}}~\enspace
Table~\ref{tbl:snr} presents the mean and standard deviation of the estimated utterance-level SNRs across all control and dysarthric utterances for the UA-Speech and TORGO databases.
As demonstrated by the large standard deviation values of the estimated SNRs, it can be said that there is a large variation in the acoustic conditions of the recorded utterances for both databases.
Most importantly, it can be observed that there is a large difference in the average SNRs of control and dysarthric utterances in both databases, with the difference being larger for the UA-Speech database\footnote{Although not presented here due to space constraints, the utterance-level SNRs have been estimated using different SNR estimators. While the absolute value of the estimated SNRs can be largely different depending on the used SNR estimator, all estimators show large standard deviation values and considerable differences between the average SNRs of control and dysarthric utterances in both databases.}.
With consistently different recording conditions between control and dysarthric utterances, there is no guarantee that automatic dysarthria classification approaches validated on these databases are learning control and dysarthric speech differences instead of differences in recording conditions for the two groups of speakers.

\begin{table}[t!]
  \def\tabcolsep{2pt}
  \scriptsize
  \begin{center}
    \caption{Mean and standard deviation of the speaker classification accuracy [$\%$] across all folds and models in the UA-Speech database.}
    \label{tbl:speaker_uas}
    \begin{tabularx}{\linewidth}{Xrrr}
      \toprule
      Approach & Speech & Non-speech & Speech\&Non-speech \\
      \midrule
      \emph{SVM+openSMILE} & $81.0 \pm 19.8$ & $ 84.5 \pm 21.9$ & $ 83.3 \pm 21.1$\\
      \emph{SVM+MFCCs} & $81.0 \pm 1.7{\color{white}{0}}$  & $ 100.0 \pm 0.0{\color{white}{0}} $ & $ 100.0 \pm 0.0{\color{white}{0}} $ \\      
      \emph{SVM+sparsity-based features} & $94.0 \pm 1.7{\color{white}{0}}  $  & $ 96.4 \pm 0.0{\color{white}{0}} $ & $96.4 \pm 0.0{\color{white}{0}} $ \\      
      \emph{CNN+Mel spectrograms} & $95.2 \pm 1.7{\color{white}{0}}  $  & $97.6 \pm 1.7{\color{white}{0}} $ & $98.8 \pm 1.7{\color{white}{0}} $ \\   
      \emph{SRL+Mel spectrograms} & $98.8 \pm 1.7{\color{white}{0}} $  & $ 100.0 \pm 0.0{\color{white}{0}}  $& $100.0 \pm 0.0{\color{white}{0}} $\\      
      \emph{MLP+ft-wav2vec2} & $ 95.2 \pm 1.7{\color{white}{0}}$  & $97.6 \pm 1.7{\color{white}{0}} $ & $95.2 \pm 1.7{\color{white}{0}}$ \\
      \emph{MLP+wav2vec2} & $ 54.8 \pm 1.7{\color{white}{0}}$  & $ 58.3 \pm 1.7{\color{white}{0}} $& $54.8 \pm 1.7{\color{white}{0}} $ \\
      \bottomrule
  \end{tabularx}
 \end{center}
\end{table}

{\emph{Dysarthria classification.}}~\enspace
Table~\ref{tbl:speaker_uas} presents the mean and standard deviation of the classification accuracy obtained on the speech segments, the non-speech segments, and on the complete utterances without using any VAD (i.e., speech\&non-speech) for the UA-Speech database using all considered approaches~(cf.~Table~\ref{tbl:summary}).
It can be observed that all approaches achieve the same or even better dysarthria classification accuracy when using non-speech segments in comparison to when using speech segments or the complete speech\&non-speech segments. 
More specifically, it can be observed that when using non-speech segments, all approaches except for {\emph{MLP+wav2vec2}} yield a high classification accuracy ranging from $84.5$\% to $100.0$\%.
The \emph{MLP+wav2vec2} approach performs considerably worse than its fine-tuned version \emph{MLP+ft-wav2vec2} and all other considered approaches. 
This result is to be expected since the representations generated by the (frozen) wav2vec2 model should be less susceptible to noise given that the model is trained on a large database of noisy speech.

\begin{table}[t!]
  \def\tabcolsep{2pt}
  \scriptsize
  \begin{center}
    \caption{Mean and standard deviation of the speaker classification accuracy [$\%$] across all folds and models in the TORGO database.}
    \label{tbl:speaker_torgo}
    \begin{tabularx}{\linewidth}{Xrrr}
      \toprule
      Approach & Speech & Non-speech & Speech\&Non-speech \\
      \midrule
      \emph{SVM+openSMILE} & $60.0 \pm 5.4{\color{white}{0}}$ & $82.2 \pm 6.3{\color{white}{0}}$ & $71.1 \pm 12.6 $\\
      \emph{SVM+MFCCs} & $60.0 \pm 0.0{\color{white}{0}}  $  & $88.9 \pm 3.1{\color{white}{0}} $ &$57.8 \pm 3.1{\color{white}{0}}$ \\      
      \emph{SVM+sparsity-based features} & $ 73.3 \pm 0.0{\color{white}{0}} $  & $ 93.3 \pm 0.0{\color{white}{0}}$ &$ 73.3 \pm 5.4{\color{white}{0}}$ \\      
      \emph{CNN+Mel spectrograms} & $53.3 \pm 11.5  $  & $77.8 \pm 10.2  $ & $68.9 \pm 10.2  $ \\      
      \emph{SRL+Mel spectrograms} & $71.1 \pm 3.1{\color{white}{0}}  $  & $ 100.0 \pm 0.0{\color{white}{0}}$& $ 91.1 \pm 3.1{\color{white}{0}} $\\      
      \emph{MLP+ft-wav2vec2} & $ 60.0 \pm 5.4{\color{white}{0}} $  & $ 57.8 \pm 3.1{\color{white}{0}} $ & $60.0 \pm 5.4{\color{white}{0}}$ \\
      \emph{MLP+wav2vec2} & $ 55.6 \pm 3.1{\color{white}{0}}$  & $ 57.8 \pm 3.1{\color{white}{0}} $& $57.8 \pm 6.3{\color{white}{0}}$ \\         
      \bottomrule
  \end{tabularx}
 \end{center}
\end{table}

Table~\ref{tbl:speaker_torgo} presents the mean and standard deviation of the classification accuracy obtained on the speech, non-speech, and the complete speech and non-speech segments from the TORGO database using all considered approaches~(cf.~Table~\ref{tbl:summary}).
Similarly to before, it can be observed that all approaches achieve the same or even better dysarthria classification accuracy when using non-speech segments in comparison to when using speech segments or the complete speech\&non-speech segments. 
Further, it can be observed that the \emph{MLP+wav2vec2} approach is not as sensitive to the recording conditions as the other approaches, as illustrated by the lower performance on non-speech segments.
However, differently from before, the performance of the fine-tuned counterpart \emph{MLP+ft-wav2vec2} on non-speech segments is also low. 
We suspect this occurs due to the much smaller amount of speech material available for fine-tuning the wav2vec2 model on the TORGO database (in contrast to the UA-Speech database).

In summary, the results presented in this section show that the majority of the considered state-of-the-art approaches achieve the same or even better dysarthria classification performance when using non-speech segments than when using speech segments or complete utterances from the UA-Speech and TORGO databases.
These results confirm our hypothesis that classification results obtained on the UA-Speech and TORGO databases can be greatly affected by characteristics of the recording environment and setup instead of dysarthria characteristics.
We hope that these databases are used with care in the future when developing and evaluating automatic dysarthric speech classification approaches.

\section{Conclusions}
In this paper, we have investigated the use of the UA-Speech and TORGO databases to validate automatic dysarthria classification approaches. 
We hypothesized that classification results obtained using these databases could be biased towards capturing characteristics of the recording environment rather than characteristics of dysarthric speech. 
To investigate this hypothesis, we have estimated the utterance-level SNRs on these databases.
Further, we have trained and validated state-of-the-art dysarthria classification approaches on the speech and non-speech segments of these databases.
Experimental results have shown that the utterance-level SNRs in control and dysarthric recordings are indeed considerably different in both databases.
Additionally confirming our hypothesis, experimental results have shown that several state-of-the-art approaches achieve the same or a considerably better dysarthria classification performance when using only the non-speech segments than when using only the speech segments of these databases.
We hope that these results raise awareness in the research community about the care that should be taken with respect to the quality of recordings when developing and evaluating automatic dysarthria classification approaches.
Further, we hope that these results foster motivation to design novel dysarthria classification approaches that are not sensitive to adverse recording conditions.

\footnotesize
\bibliographystyle{IEEEbib}
\bibliography{refs}

\end{document}